\documentclass[11pt]{article}
\usepackage {amsfonts}
\usepackage {graphicx}
\usepackage {epsfig}
\usepackage{amssymb}
\usepackage{amsmath}
\usepackage {longtable}
\usepackage[english]{babel}
\begin{document}

\title{The deuteron (nuclei) birefringence effect in a matter and
in an electric field and the searches for an EDM of a deuteron
(nucleus) rotating in a storage ring}

\author{V.G. Baryshevsky, A.R. Shirvel \\ Research Institute for Nuclear Problems, Belarusian State
University,\\ 11 Bobruyskaya Str., Minsk 220050, Belarus,
\\ e-mail: bar@inp.minsk.by}

\maketitle

\begin{center}
\begin{abstract}
The phenomena of deuteron
 birefringence in a matter and an electric field should be
 accurately considered when preparing experiments for the EDM
 search with a storage ring, because they could imitate the spin
 rotation due to the EDM. Moreover, study of these effects in such
 experiments could
 provide to measure both the spin-dependent part of the amplitude of
 the coherent elastic scattering of a deuteron by a nucleus at the
 zero angle and the tensor electric polarizability of a deuteron.
\end{abstract}
\end{center}

 \section{INTRODUCTION}

The phenomena of spin rotation and spin dichroism (birefringence
effect) for particles with the spin $S\geq1$ in an unpolarized
medium were theoretically described for the first time in
\cite{1,2}.
 Deuteron spin dichroism was observed for the first time with the $20$ MeV
accelerator \cite{3}.
 Further investigations of this phenomenon are planned to be carry
out with a storage ring and an external beam \cite{4,5}.
{ Observation of particle spin rotation and spin dichroism}
(birefringence effect) with a storage ring requires reducing of
$(g-2)$ precession frequency ($g$ is the gyromagnetic ratio).
 This precession appears due to interaction of the particle magnetic moment
 with an external electromagnetic field.
 The requirement for $(g-2)$ precession cancellation also arises when searching for a deuteron electric
dipole moment (EDM) with a storage ring by the deuteron spin
precession in an electric field \cite{6, project}.
 According to \cite{6} balancing the energy of the
particle and the strength of the electric field in a storage ring
provides to reduce and even zeroize the $(g-2)$ precession
frequency.
As a result, the EDM-caused spin rotation grows linearly with time
\cite{6,project}.
 Note that, when $(g-2)$ precession is suppressed, the angle of spin rotation induced by the birefringence effect
 grows linearly with time, too.

 The effect of deuteron (nucleus) birefringence in a medium reveals itself in a
storage ring due to presence of the residual gas inside the
storage ring and use of a gas jet (gas target) for deuteron
(nucleus) polarization analysis.
 Moreover, the birefringence also occurs in a
solid target used for analysis of polarization of the deuteron
(nucleus) beam outside the storage ring.
Therefore, the phenomenon of birefringence in the gas medium and
polarimeter would appear as a systematic error in the EDM
measurements \cite{project}.
%
 In addition, study of the birefringence phenomenon is of self-importance
since it makes possible to measure the spin-depended part of the
forward scattering amplitude.

Lastly, the action of the electric field on the deuteron rouses
one more mechanism of deuteron spin rotation and oscillations (the
phenomenon of birefringence in an electric field)
 conditioned by the deuteron
tensor electric polarizability \cite{7}.

In this paper the deuteron birefringence in a matter and in an
electric field is considered for a particle moving in a storage
ring.
 The equations describing the behavior of the deuteron spin in a storage ring
including all the above mentioned contributions are derived.
 It is
shown that the birefringence effect is noticeable and should be
considered when carrying out the deuteron (nucleus) EDM searches
at a storage ring.

\section{THE PHENOMENON OF BIREFRINGENCE}

According to the analysis \cite{1,2}, when a particle with the
spin $S\geq1$ passes through an unpolarized medium, the medium
refraction index depends on the particle spin orientation to its
momentum. Therefore, the particle possesses some effective
potential energy $V$ in the medium and this energy depends on the
spin orientation \cite{1,2,4}
\begin{equation}
\hat{V}=-\frac{2\pi \hbar^{2}}{M \gamma}N\hat{f(0)},
 \label{1.1}
\end{equation}
 where $M$ is the particle mass, $\hat{f(0)}$ is the
spin dependent zero-angle elastic coherent scattering amplitude of
the particle, $N$ is the density of the scatterers in the matter
(the number of scatterers in $1cm^{3}$), $\gamma$ is the Lorentz
factor.
 Substituting $\hat{f(0)}$ for a particle with the spin
 $S=1$  in (\ref{1.1}) in the explicit form one can obtain \cite{1,2,4}
\begin{equation}
\hat{V}=-\frac{2\pi
\hbar^{2}}{M\gamma}N\left(d+d_{1}\left(\vec{S}\vec{n}\right)^{2}\right),
\label{1.2}
\end{equation}
 where  $\vec{n}$ is the unit vector along the particle momentum
direction.

Let the quantization axis z is directed along $\vec{n}$ and $m$
denotes the magnetic quantum number. Then, for a particle in a
state that is an eigenstate of the operator $S_{z}$ of spin
projection onto the z-axis, the efficient potential energy can be
written as:
\begin{equation}
\hat{V}=-\frac{2\pi \hbar^{2}}{M\gamma}N\left(d+d_{1}m^{2}\right).
\label{1.3}
\end{equation}

 According to (\ref{1.3}) splitting of the deuteron energy levels
in a matter is similar to splitting of atom energy levels in an
electric field aroused by the quadratic Stark effect.
 Therefore,
the above effect could be considered as caused by splitting of the
spin levels of the particle in the pseudoelectric nuclear field of
a matter.

Let a real electric field $\vec{E}$ acts on a deuteron (nucleus).
The energy $\hat{V}_{E}$ of deuteron beam in an external electric
field due to the tensor electric polarizability can be written in
the form
\begin{equation}
\hat{V}_{E}=-\frac{1}{2}\hat{\alpha}_{ik}E_{i}E_{k},
 \label{1.4}
\end{equation}
where $\hat{\alpha}_{ik}$ is the deuteron tensor electric
polarizability, $E_{i}$ are the components of the electric field.
This expression can be rewritten as follows:
\begin{equation}
\hat{V}_{E}=\alpha_{S}E^{2}-\alpha_{T}E^{2}\left(\vec{S}\vec{n}_{E}\right)^{2},
 \label{1.5}
\end{equation}
where $\alpha_{S}$ is the deuteron scalar electric polarizability,
$\alpha_{T}$ is the deuteron tensor electric polarizability,
$\vec{n}_{E}$ is the unit vector along $\vec{E}$.

 {Comparing (\ref{1.5}) with (\ref{1.2}) we can conclude that the
effect of spin rotation and oscillations about the $\vec{E}$
direction can be observed for particle with $S \ge 1$ in an
electric field, too \cite{7}.}

 Thus, considering evolution of the spin of a particle in a
storage ring one should take into account several interactions:

1. interactions of magnetic and electric dipole moments with an
electromagnetic field;

2. interaction (\ref{1.5}) of a particle with an electric field
due to the tensor electric polarizability

3. interaction (\ref{1.2}) of a particle with the pseudoelectric
nuclear field of a matter.

Therefore, the equation for the particle spin wavefunction is:
\begin{equation}
i\hbar\frac{\partial\Psi(t)}{\partial
t}=\left(\hat{H}_{0}+\hat{V}_{d}+\hat{V}+\hat{V}_{E}\right)\Psi(t)
\label{1.6}
\end{equation}
where $\Psi(t)$ is the particle spin wavefunction,
{$\hat{H}_{0}$ is the Hamiltonian describing the spin behavior
caused by interaction of the magnetic moment with the
electromagnetic field (equation (\ref{1.6}) with the only
$\hat{H}_{0}$  summand converts to the Bargman-Myshel-Telegdy
equation),}
 $\hat{V}_{d}$ describes interaction of the deuteron (nuclear)
EDM with an electric field.

\section{THE EQUATIONS FOR THE POLARIZATION VECTOR AND
QUADRUPOLARIZATION TENSOR OF THE DEUTERON BEAM IN A STORAGE RING}

Let us consider motion of a deuteron in a storage ring in external
magnetic and electric fields. Particle spin precession induced by
interaction of the magnetic moment of a particle with an external
electromagnetic field  can be described by the
Bargman-Myshel-Telegdy equation \cite{6,8}
\begin{equation}
\frac{d\vec{p}}{dt}=[\vec{p}\times\vec{\Omega}_{0}],
\label{2.1}
\end{equation}
where $t$ is time in the laboratory system,
\begin{equation}
\vec{\Omega}_{0}=\frac{e}{mc}\left[\left(a+\frac{1}{\gamma}\right)\vec{B}
-a\frac{\gamma}{\gamma+1}\left(\vec{\beta}\cdot\vec{B}\right)\vec{\beta}-
\left(\frac{g}{2}-\frac{\gamma}{\gamma+1}\right)\vec{\beta}\times\vec{E}\right],
\label{2.2}
\end{equation}
$m$ is the mass of the particle, $e$ is its charge,$\vec{p}$ is
the spin polarization vector , $\gamma$ is the Lorentz-factor,
 $\vec{\beta}=\vec{v}/c$, $\vec{v}$ is the particle velocity, $a=(g-2)/2$, $g$ is the gyromagnetic ratio, $\vec{E}$ and
 $\vec{H}$ are the electric and magnetic fields in the point of
 particle location.

If a particle possesses an intrinsic dipole moment then the
additional term that describes the spin rotation induced by the
EDM should be added to (\ref{2.1}) \cite{6}
\begin{equation}
\frac{d\vec{p}_{edm}}{dt}=\frac{d}{\hbar}
\left[\vec{p}\times\left(\vec{\beta}\times\vec{B}+\vec{E}\right)\right],
\label{2.3}
\end{equation}
where $d$ is the electric dipole moment of a particle.

As a result, evolution of the deuteron spin due to the magnetic
and electric momenta can be described by the following equation:
\begin{eqnarray}
\frac{d\vec{p}}{dt}=
\frac{e}{mc}\left[\vec{p}\times\left\{\left(a+\frac{1}{\gamma}\right)\vec{B}
-a\frac{\gamma}{\gamma+1}\left(\vec{\beta}\cdot\vec{B}\right)\vec{\beta}-
\left(\frac{g}{2}-\frac{\gamma}{\gamma+1}\right)\vec{\beta}\times\vec{E}\right\}\right]
+d\left[\vec{p}\times\left(c\vec{\beta}\times\vec{B}+\vec{E}\right)\right].
\label{2.4}
\end{eqnarray}

According to the section 2, the equation (\ref{2.4}) does not
describe particle spin evolution in a storage ring completely.
 The expression (\ref{2.4}) should be
supplemented with the additions given by interactions
$\hat{V}_{E}$ and $\hat{V}$ (see (\ref{1.2}-\ref{1.6})) .

This additional contribution could be found by the aids of the
particle spin wavefunction $\Psi(t)$ (see \ref{1.6})).

 The equations describing the time evolution of the spin
and tensor of quadrupolarization {caused by the phenomena of
birefringence} can be written as:
\begin{eqnarray}
\frac{d\vec{p}}{dt}& = &
\frac{d}{dt}\frac{\langle\Psi(t)|\vec{S}|\Psi(t)\rangle}{\langle\Psi(t)|\Psi(t)|},
\nonumber \\
\frac{dp_{ik}}{dt} & = &
\frac{d}{dt}\frac{\langle\Psi(t)|Q_{ik}|\Psi(t)\rangle}{\langle\Psi(t)|\Psi(t)|},
\label{2.5}
\end{eqnarray}
where $\Psi(t)$ is the deuteron wave function,
$\hat{Q}_{ik}=\frac{3}{2}\left(S_{i}S_{k}+S_{k}S_{i}-\frac{4}{3}\delta_{ik}\hat{\texttt{I}}\right)$
is the tensor of rank two (tensor of quadrupolarization).

 The equations (\ref{2.5}) contain initial phases that
determine the deuteron wave function.
 Therefore, a partly
polarized beam can not be described by such equations.
 So the
 spin density matrix formalism should be used to derive
equations describing the evolution of the deuteron spin.

The density matrix of the system "deuteron+target" is
\begin{eqnarray}
\rho=\rho_{d}\otimes\rho_{t},
 \label{2.6}
\end{eqnarray}
where $\rho_{d}$ is the density matrix of the deuteron beam
\begin{eqnarray}
\rho_{d}=I(\vec{k})\left(\frac{1}{3}\hat{\texttt{I}}
+\frac{1}{2}\vec{p}(\vec{k})\vec{S}+\frac{1}{9}p_{ik}(\vec{k})\hat{Q}_{ik}\right),
\label{2.7}
\end{eqnarray}
$I(\vec{k})$ is the intensity of the beam, $\vec{p}$ is the
polarization vector, $p_{ik}$ is the quadrupolarization tensor of
the deuteron beam, $\rho_{t}$ is the density matrix of the target.
{For an unpolarized target $\rho_{t}=\hat{\texttt{I}}$, where $I$
is the unit matrix in the spin space of target particle.}

The equation for the deuteron beam density matrix can be written
as:
\begin{eqnarray}
\frac{d\rho_{d}}{dt}=-\frac{i}{\hbar}\left[\hat{H},\rho_{d}\right]+\left(\frac{\partial\rho_{d}}{\partial
t}\right)_{col},
\label{2.9}
\end{eqnarray}
where $\hat{H}=\hat{H}_{0}+\hat{V}_{d}+\hat{V}_{E}$,
\begin{eqnarray}
\hat{V}_{d} & = &
-d\left(\vec{\beta}\times\vec{B}+\vec{E}\right)\vec{S}
\label{2.10}\\
\hat{V}_{E} & = &
\alpha_{S}\left(\vec{\beta}\times\vec{B}+\vec{E}\right)^{2}
-\alpha_{T}\left(\vec{\beta}\times\vec{B}+\vec{E}\right)^{2}\left(\vec{S}\vec{n}_{E}\right)^{2},
\nonumber \\
\vec{n}_{E} & = &
\frac{\vec{E}+\vec{\beta}\times\vec{B}}{|\vec{E}+\vec{\beta}\times\vec{B}|}.
\nonumber
\end{eqnarray}
 The collision term
$\left(\frac{\partial\rho_{d}}{\partial t}\right)_{col}$ can be
found by the method described in \cite{9}
\begin{eqnarray}
\left(\frac{\partial\rho_{d}}{\partial
t}\right)_{col}
=vN\emph{Sp}_{t}\left[\frac{2\pi
i}{k}\left[F(\theta=0)\rho-\rho F^{+}(\theta=0)\right] +\int
d\Omega F(\vec{k}^{'})\rho(\vec{k}^{'})F^{+}(\vec{k}^{'})\right],
\label{2.11}
\end{eqnarray}
where $\vec{k}^{'}=\vec{k}+\vec{q}$, {$\vec{q}$ is the momentum
carried over a nucleus of the matter from the incident particle,}
 $v$ is the speed of the incident particles, $N$ is the
atom density in the matter,
 $F$ is the scattering amplitude
depending on the spin operators of the deuteron and the matter
nucleus (atom), $F^+$ is the Hermitian conjugate of the operator
$F$.
{The first term in (\ref{2.11}) describes coherent scattering of a
particle by matter nuclei, while the second term is for multiple
scattering.}

Let us consider the first term in (\ref{2.11}):
\begin{eqnarray}
\left(\frac{\partial\rho_{d}}{\partial
t}\right)_{col}^{(1)}=vN\frac{2\pi i}{k}
 \left[
 \hat{f}(0)\rho_d-\rho_d \hat{f}(0)^{+}
\right] .
\label{2.12}
\end{eqnarray}
The  amplitude $\hat{f}(0)$ of deuteron scattering in an
unpolarized target at the zero angle is
\begin{eqnarray}
\hat{f}(0)=\emph{Sp}_{t}F(0)\rho_{t}.
\label{2.13}
\end{eqnarray}
This amplitude can de rewritten according (\ref{1.2}) as
\begin{eqnarray}
\hat{f}(0)=d+d_{1}(\vec{S}\vec{n})^{2},
\label{2.14}
\end{eqnarray}
where $\vec{n}=\vec{k}/k$, $\vec{k}$ is the deuteron momentum.

As a result one can obtain:
\begin{eqnarray}
\left(\frac{\partial\rho_{d}}{\partial t}\right)_{col}^{(1)} =
 -\frac
i\hbar\left(\hat{V} {\rho_d}-{\rho_d} \hat{V}^{+}\right).
\label{2.15}
\end{eqnarray}

Finally, the expression (\ref{2.9}) reads
\begin{eqnarray}
\frac{d\rho_{d}}{dt}=-\frac{i}{\hbar}\left[\hat{H},\rho_{d}\right]
 -\frac
i\hbar\left(\hat{V} {\rho_d}-{\rho_d} \hat{V}^{+}\right)+
 vN
\emph{Sp}_{t} \int d\Omega
F(\vec{k}^{'})\rho(\vec{k}^{'})F^{+}(\vec{k}^{'}).
 \label{2.9_new}
\end{eqnarray}
The last term in the above formula, which is proportional to
$\emph{Sp}_{t}$, describes the multiple scattering process and
spin depolarization aroused from it. Henceforward we consider such
time of experiment (such effective length for a particle in a
matter) that provides to neglect this term.

The intensity of the beam is
\begin{eqnarray}
I=\emph{Sp}_{d}\rho_{d}.
\label{2.16}
\end{eqnarray}
Consequently
\begin{eqnarray}
\frac{dI}{dt}=vN\frac{2\pi
i}{k}\emph{Sp}_{d}\left[f(0)\rho_{d}-\rho_{d}f^{+}(0)\right].
\label{2.17}
\end{eqnarray}
Substituting (\ref{2.7}) and (\ref{2.14}) into (\ref{2.17}) we can
get
\begin{eqnarray}
\frac{dI}{dt}=\frac{\chi}{3}\left[2+p_{ik}n_{i}n_{k}\right]I(t)+\alpha
I(t),
\label{2.18}
\end{eqnarray}
where $\chi=-\frac{4\pi
vN}{k}\texttt{Im}d_{1}=--vN(\sigma_1-\sigma_0)$,
$\alpha=-\frac{4\pi vN}{k}\texttt{Im}d=-v N \sigma_0$, $\sigma_1$
and $\sigma_0$ are the total cross-sections of deuteron scattering
by a nonpolarized nucleus for the magnetic quantum numbers $m=1$
and $m=0$, respectively.

 Polarization vector of the
deuteron beam $\vec{p}$ is determined as
\begin{eqnarray}
\vec{p}=\frac{\emph{Sp}_{d}\rho_{d}\vec{S}}{\emph{Sp}_{d}\rho_{d}}=\frac{\emph{Sp}_{d}\rho_{d}\vec{S}}{I}.
\label{2.19}
\end{eqnarray}

From (\ref{2.19}) one can get the differential equation for the
beam polarization
\begin{eqnarray}
\frac{d\vec{p}}{dt}=\frac{\emph{Sp}_{d}(d\rho_{d}/dt)\vec{S}}{I(t)}-
\vec{p}\frac{\emph{Sp}_{d} (d\rho_{d}/dt)}{I(t)}.
\label{2.20}
\end{eqnarray}

The expression for the quadrupolarization tensor is
\begin{eqnarray}
p_{ik}=\frac{\emph{Sp}_{d}\rho_{d}Q_{ik}}{\emph{Sp}_{d}\rho_{d}}=\frac{\emph{Sp}_{d}\rho_{d}Q_{ik}}{I},
\label{2.21}
\end{eqnarray}
where
$\hat{Q}_{ik}=\frac{3}{2}\left(S_{i}S_{k}+S_{k}S_{i}-\frac{4}{3}\delta_{ik}\hat{\texttt{I}}\right)$.

The change of the quadrupolarization tensor can be written as
\begin{eqnarray}
\frac{dp_{ik}}{dt}=\frac{\emph{Sp}_{d}(d\rho_{d}/dt)Q_{ik}}{I(t)}-
p_{ik}\frac{\emph{Sp}_{d} (d\rho_{d}/dt)}{I(t)}.
\label{2.22}
\end{eqnarray}

Using (\ref{2.7}) and (\ref{2.1}), (\ref{2.20}) and (\ref{2.22})
we can get the equation system  for the time evolution of the
deuteron polarization vector and quadrupolarization tensor
($\vec{n}=\vec{k}/k$,
$\vec{n}_{E}=\frac{\vec{E}+\vec{\beta}\times\vec{B}}{|\vec{E}+\vec{\beta}\times\vec{B}|}$,
$p_{xx}+p_{yy}+p_{zz}=0$)
\begin{eqnarray}
\left\{
\begin{array}{l}
\frac{d\vec{p}}{dt}=
\frac{e}{mc}\left[\vec{p}\times\left\{\left(a+\frac{1}{\gamma}\right)\vec{B}
-a\frac{\gamma}{\gamma+1}\left(\vec{\beta}\cdot\vec{B}\right)\vec{\beta}-
\left(\frac{g}{2}-\frac{\gamma}{\gamma+1}\right)\vec{\beta}\times\vec{E}\right\}\right]+\\
 +
\frac{d}{\hbar}\left[\vec{p}\times\left({\vec{E}}+\vec{\beta}\times\vec{B}\right)\right]
+\frac{\chi}{2}(\vec{n}(\vec{n}\cdot\vec{p})+\vec{p}) + \\
 +  \frac{\eta}{3}[\vec{n}\times\vec{n}^{'}]
-\frac{2\chi}{3}\vec{p}-\frac{\chi}{3}(\vec{n}\cdot\vec{n}^{'})\vec{p}
-\frac{2}{3}\frac{\alpha_{T}E^{2}}{\hbar}[\vec{n}_{E}\times\vec{n}_{E}^{'}],\\
{} \\
\frac{dp_{ik}}{dt}  =  -\left(\varepsilon_{jkr}p_{ij}\Omega_{r}+\varepsilon_{jir}p_{kj}\Omega_{r}\right) + \\
 +
\chi\left\{-\frac{1}{3}+n_{i}n_{k}+\frac{1}{3}p_{ik}-\frac{1}{2}(n_{i}^{'}n_{k}+n_{i}n_{k}^{'})
+\frac{1}{3}(\vec{n}\cdot\vec{n}^{'})\delta_{ik}\right\} + \\
 +
\frac{3\eta}{4}\left([\vec{n}\times\vec{p}]_{i}n_{k}+n_{i}[\vec{n}\times\vec{p}]_{k}\right)
-\frac{\chi}{3}(\vec{n}\cdot\vec{n}^{'})p_{ik} - \\
 -
\frac{3}{2}\frac{\alpha_{T}E^{2}}{\hbar}\left([\vec{n}_{E}\times\vec{p}]_{i}n_{E,\,k}
+n_{E,\,i}[\vec{n}_{E}\times\vec{p}]_{k}\right),
\\
\end{array}
\right.
\label{2.23}
\end{eqnarray}
where $\eta=-\frac{4 \pi N}{k} \texttt{Re}d_{1}$,
$n_{i}^{'}=p_{ik}n_{k}$, $n_{E,\,i}^{'}=p_{ik}n_{E,\,k}$\,,
$\Omega_{r}(d)$ are the components of the vector $\vec{\Omega}(d)$
($r=1,2,3$ corresponds $x,y,z$):
\begin{eqnarray}
\vec{\Omega}(d) & = &
\frac{e}{mc}\left\{\left(a+\frac{1}{\gamma}\right)\vec{B}
-a\frac{\gamma}{\gamma+1}\left(\vec{\beta}\cdot\vec{B}\right)\vec{\beta}-
\left(\frac{g}{2}-\frac{\gamma}{\gamma+1}\right)\vec{\beta}\times\vec{E}\right\} + \nonumber \\
& + &
\frac{d}{\hbar}\left({\vec{E}}+\vec{\beta}\times\vec{B}\right).
\label{2.24}
\end{eqnarray}

Then we consider the spin rotation about the particle momentum.
 According to \cite{6,project} the spin precession caused by the magnetic
moment ($(g-2)$ precession) can be minimized and even zeroized by
applying a radial electric field.

{The angles of spin rotation caused by both the EDM and
birefringence effect are small for the considered experiment
duration.}
 Therefore, the perturbation theory can be used for (\ref{2.23})
 solution.
\begin{eqnarray}
\vec{p}(t)=\vec{p}^{\,0}+\frac{e}{mc}\left[\vec{p}^{\,0}\times\left\{a\vec{B}
+\left(\frac{1}{\gamma^{2}-1}-a\right)\vec{\beta}\times\vec{E}\right\}\right]t
+
\nonumber \\
+\frac{d}{\hbar}\left[\vec{p}^{\,0}\times\left({\vec{E}}+\vec{\beta}\times\vec{B}\right)\right]t
+ \{
\frac{\chi}{2}(\vec{n}(\vec{n}\cdot\vec{p}^{\,0})+\vec{p}^{\,0})t
+\frac{\eta}{3}[\vec{n}\times\vec{n}^{'}_{0}]t - \nonumber \\
-\frac{2\chi}{3}\vec{p}^{0}t
-\frac{\chi}{3}(\vec{n}\cdot\vec{n}^{'}_{0})\vec{p}^{\,0}t \}
-\frac{2}{3}\frac{\alpha_{T}E^{2}}{\hbar}[\vec{n}_{E}\times\vec{n}_{E\,0}^{'}]t,
\label{2.25}
\end{eqnarray}
\begin{eqnarray}
\lefteqn
p_{ik}(t)  =
p_{ik}^{0}-\frac{e}{mc}(\varepsilon_{jkr}p_{ij}+\varepsilon_{jir}p_{kj})\left\{a\vec{B}
+\left(\frac{1}{\gamma^{2}-1}-a\right)\vec{\beta}\times\vec{E}\right\}_{r}t - \nonumber\\
-
\frac{d}{\hbar}(\varepsilon_{jkr}p_{ij}+\varepsilon_{jir}p_{kj})\left({\vec{E}}
+\vec{\beta}\times\vec{B}\right)_{r}t + \nonumber \\
+ \{
 \chi\left[-\frac{1}{3}+n_{i}n_{k}+\frac{1}{3}p_{ik}^{0}
-\frac{1}{2}(n_{i0}^{'}n_{k}+n_{i}n_{k0}^{'})
+\frac{1}{3}(\vec{n}\cdot\vec{n}^{'}_{0})\delta_{ik}\right]t + \nonumber \\
 +  \frac{3\eta}{4}\left([\vec{n}\times\vec{p}^{\,0}]_{i}n_{k}
+n_{i}[\vec{n}\times\vec{p}^{\,0}]_{k}\right)t
-\frac{\chi}{3}(\vec{n}\cdot\vec{n}^{'}_{0})p_{ik}^{\,0}t \}-
\nonumber \\
-\frac{3}{2}\frac{\alpha_{T}E^{2}}{\hbar}\left([\vec{n}_{E}\times\vec{p}^{\,0}]_{i}n_{E,\,k}
+n_{E,\,i}[\vec{n}_{E}\times\vec{p}^{\,0}]_{k}\right)t,
\label{2.26}
\end{eqnarray}
where $\vec{p}^{\,0}$ is the beam polarization at $t_{0}=0$,
$n_{i0}^{'}=p_{ik}^{\,0}n_{k}$,
$n_{E\,0,\,i}^{'}=p_{ik}^{\,0}n_{E,\,k}$, $p_{ik}^{\,0}$ is the
components of the quadrupolarization tensor at the initial moment
of time.

In real situation, even when $(g-2)$ precession is suppressed,
nevertheless, the rotation angle can appear large enough (during
the experiment the spin can rotate several turns \cite{project}).
Absorption can also appear significant. In this case one should
analyze the system (\ref{2.23}) instead of perturbation theory
results (\ref{2.25}).

Thus, according to (\ref{2.25}), (\ref{2.26}) the  spin behavior
of a deuteron rotating in the storage ring is caused by several
contributions:

1. spin rotation which is described by Bargman-Myshel-Telegdy
equation;

2. rotation due to the deuteron EDM,

3.rotation and dichroism due to the the phenomena of birefringence
in a medium and

4. spin rotation due to the phenomena of birefringence in an
electric field.

Let us consider some particular cases.

\textbf{Case I.}
Suppose the vector polarization is parallel to the the $z$-axis,
i.e. $p_{x}^{\,0}=p_{y}^{\,0}=0$, $p_{z}^{\,0}\neq0$,
$p_{ik}^{\,0}=0$, if  $i\neq k$,  $p_{xx}\neq 0$,
$p_{yy}^{\,0}\neq 0$, $p_{zz}^{\,0}=0$.

\begin{figure}[h]
\begin{center}
\begin{picture}(40,40)\thicklines
\put(0,0){\vector(1,0){40}} \put(0,0){\vector(0,1){40}}
\put(0,0){\vector(3,4){20}} \put(38,2){z} \put(4,34){y}
\put(22,20){x} \put(0,0){\vector(1,0){25}} \put(16,-10){$\vec{p}$}
\end{picture}
\end{center}
\caption{The initial orientation of the polarization
vector}
\label{Fig.1}
\end{figure}
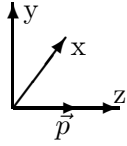

Suppose $(g-2)$ spin precession caused by the magnetic moment is
zero, then one can obtain
\begin{eqnarray}
p_{x}(t)=0 \nonumber\\
p_{y}(t)=-\frac{d}{\hbar}\,p_{z}^{\,0}\left({E}+\beta
B\right)t , \nonumber\\
p_{z}(t)=p_{z}^{\,0}+\frac{1}{3}\chi p_{z}^{\,0}t , \nonumber\\
p_{xx}(t)=p_{xx}^{\,0}+\frac{\chi}{3}\left(-1+p_{xx}^{\,0}\right)t , \nonumber\\
p_{yy}(t)=p_{yy}^{\,0}+\frac{\chi}{3}\left(-1+p_{yy}^{\,0}\right)t , \\
p_{zz}(t)=\frac{2\chi}{3}t , \nonumber\\
p_{xy}(t)=p_{xz}(t)=0 , \nonumber\\
p_{yz}(t)=\frac d \hbar \,p_{yy}\left(E +\beta B\right)t, \nonumber\\
p_{xy}=\frac{3}{2}\frac{\alpha_{T}E^{2}}{\hbar}p_{z}^{\,0}
\nonumber
\label{2.27}
\end{eqnarray}
The solution (\ref{2.27}) shows that
 even in this practically ideal case (when the polarization vector is exactly parallel to $\vec{n}$)
 change of polarization due to birefringence effect leads to the
 appearance of additional components of $\vec{p}$ and $p_{ik}$
 along with the components aroused
 by the deuteron EDM.

 Case II.
 Let us consider now the more real case.

Suppose the angle between the initial polarization vector and the
$z$ axis is acute

\begin{figure}[h]
\begin{center}
\begin{picture}(40,40)\thicklines
\put(0,0){\vector(1,0){40}} \put(0,0){\vector(0,1){40}}
\put(0,0){\vector(3,4){20}} \put(38,2){z} \put(4,34){y}
\put(22,20){x} \put(0,0){\vector(4,1){28}} \put(24,8){$\vec{p}$}
\end{picture}
\end{center}
\caption{The initial orientation of the polarization vector}
\label{Fig.2}
\end{figure}
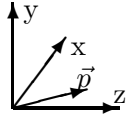
then the solution of (\ref{2.24},\ref{2.25}) can be written as
follows:
\begin{eqnarray}
p_{x}(t)=p_{x}^{\,0}-\frac{\chi}{6}\,p_{x}^{\,0}\,t-\frac{\chi}{3}\,p_{zz}^{\,0}\,p_{x}^{\,0}\,t
-\frac{\eta}{3}\,p_{yz}^{\,0}t \nonumber \\
p_{y}(t)=p_{y}^{\,0}-\frac{d}{\hbar}\left({E}+\beta
B\right)p_{z}^{\,0}\,t-\frac{\chi}{6}\,p_{y}^{\,0}\,t
-\frac{\chi}{3}\,p_{zz}^{\,0}\,p_{y}^{\,0}\,t+\frac{\eta}{3}\,p_{xz}^{\,0}t \nonumber \\
p_{z}(t)=p_{z}^{\,0}+\frac{d}{\hbar}\left({E}+\beta
B\right)p_{y}^{\,0}\,t+\frac{\chi}{3}\,p_{z}^{\,0}\,t
-\frac{\chi}{3}\,p_{zz}^{\,0}\,p_{z}^{\,0}\,t\nonumber \\
p_{xx}(t)=p_{xx}^{\,0}-\frac{\chi}{3}\left(1+p_{yy}^{\,0}\right)\,t
-\frac{\chi}{3}\,p_{zz}^{\,0}\,p_{xx}^{\,0}\,t\nonumber \\
p_{yy}(t)=p_{yy}^{\,0}-2\frac d\hbar\,p_{yz}\left(E+\beta
B\right)-\frac{\chi}{3}\left(1+p_{xx}^{\,0}\right)\,t
-\frac{\chi}{3}\,p_{zz}^{\,0}\,p_{yy}^{\,0}\,t \\
p_{zz}(t)=p_{zz}^{\,0}+2\frac{d}\hbar\,p_{yz}\left(E+\beta B\right)+\frac{\chi}{3}\left(2-p_{zz}^{\,0}\right)\,t-\frac{\chi}{3}\,p_{zz}^{\,0\,2}\,t\nonumber \\
p_{xy}(t)=p_{xy}^{\,0}-\frac{d}\hbar\,p_{xz}\left(E+\beta B\right)+\frac{\chi}{3}\,p_{xy}^{\,0}\,t-\frac{\chi}{3}\,p_{zz}^{\,0}\,p_{xy}^{\,0}\,t\nonumber \\
p_{xz}(t)=p_{xz}^{\,0}+\frac{d}\hbar\,p_{xy}\left(E+\beta
B\right)-\frac{\chi}{6}\,p_{xz}^{\,0}\,t-\frac{3\eta}{4}\,p_{y}^{\,0}\,t
-\frac{\chi}{3}\,p_{zz}^{\,0}\,p_{xz}^{\,0}\,t\nonumber \\
p_{yz}(t)=p_{yz}^{\,0}+\frac{d}\hbar\,(p_{yy}-p_{zz})\left(E+\beta
B\right)-\frac{\chi}{6}\,p_{yz}^{\,0}\,t+\frac{3\eta}{4}\,p_{x}^{\,0}\,t
-\frac{\chi}{3}\,p_{zz}^{\,0}\,p_{yz}^{\,0}\,t. \nonumber
\label{2.28}
\end{eqnarray}

According to (\ref{2.28})
 the change in components of polarization vector and tensor of
 quadrupolarization caused by the EDM is mixed with the
 contributions from the birefringence effect to the same
 components.

Thus, the changes in the deuteron polarization vector
 and quadrupolarization tensor are the result of several mechanisms:

- the rotation of the spin in the horizontal plain
$(\vec{E},\vec{n})$.

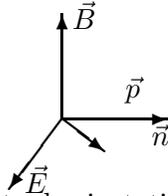
\begin{figure}[h]
\begin{center}
\begin{picture}(40,40)\thicklines
 \put(0,0){\vector(1,0){40}} \put(0,0){\vector(0,1){40}}
 \put(0,0){\vector(-3,-4){20}} \put(34,-10){$\vec{n}$}
 \put(4,34){$\vec{B}$} \put(-14,-28){$\vec{E}$}
 \put(0,0){\vector(4,-3){16}}
 \put(24,8){$\vec{p}$}
\end{picture}
\end{center}
\caption{Mutual orientation of $\vec{E}$, $\vec{B}$, and
$\vec{n}$} \label{Fig.4}
\end{figure}

This rotation is risen due to interaction of the magnetic moment
with external fields. The rotation frequency is expressed as:
\begin{equation}
\vec{\omega}_{a}=\frac{e}{mc}\left\{a\vec{B}+\left(\frac{1}{\gamma^{2}-1}
-a\right)\vec{\beta}\times\vec{E}\right\};
\label{2.29}
\end{equation}

- the rotation of the spin in the vertical plane
$(\vec{B},\vec{n})$  caused by the electric dipole moment;

\begin{figure}[ht]
\begin{center}
\begin{picture}(40,40)\thicklines
\put(0,0){\vector(1,0){40}}
 \put(0,0){\vector(0,1){40}}
 \put(0,0){\vector(-3,-4){20}}
 \put(34,-10){$\vec{n}$}
 \put(4,34){$\vec{B}$}
 \put(-14,-28){$\vec{E}$}
 \put(0,0){\vector(4,-3){16}}
 \put(24,8){$\vec{p}$}
 \put(0,0){\circle{100}}
\end{picture}
\end{center}
\caption{Rotation of the polarization vector due to birefringence
in an electric field}
\label{Fig.5}
\end{figure}

The rotation frequency is
\begin{equation}
\vec{\omega}_{d}=\frac{d}{\hbar}\left(E+\vec{\beta}\times\vec{B}\right)
\label{2.30}
\end{equation}

- the rotation caused by the phenomenon of birefringence in a
medium, this is precession in the vertical plane
$(\vec{B},\vec{E})$;

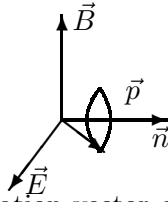
\begin{figure}[ht]
\begin{center}
\begin{picture}(40,40)\thicklines
\put(0,0){\vector(1,0){40}}
 \put(0,0){\vector(0,1){40}}
 \put(0,0){\vector(-3,-4){20}}
 \put(34,-10){$\vec{n}$}
 \put(4,34){$\vec{B}$}
 \put(-14,-28){$\vec{E}$}
 \put(0,0){\vector(4,-3){16}}
 \put(24,8){$\vec{p}$} \
 \qbezier(14,-12)(4,0)(14,12)
 \qbezier(14,-12)(22,0)(14,12)
\end{picture}
\end{center}
\caption{Rotation of the polarization vector due to the phenomena
of birefringence} \label{Fig.6}
\end{figure}

The rotation frequency is
\begin{equation}
\omega=\frac{2\pi N}{M
\gamma}\,\hbar\,\texttt{Re}d_{1}.
\label{2.31}
\end{equation}

-the rotation due to the phenomenon of birefringence in an
electric field in the vertical plane $(\vec{B},\vec{n})$, i.e. in
the same plane as the rotation caused by the EDM.

\begin{figure}[ht]
\begin{center}
\begin{picture}(40,40)\thicklines
\put(0,0){\vector(1,0){40}} \put(0,0){\vector(0,1){40}}
\put(0,0){\vector(-3,-4){20}} \put(34,-10){$\vec{n}$}
\put(4,34){$\vec{B}$} \put(-14,-28){$\vec{E}$}
\put(0,0){\vector(4,-3){16}} \put(24,8){$\vec{p}$}
\put(0,0){\circle{100}}
\end{picture}
\end{center}
\caption{Rotation of the polarization vector due to birefringence
in an electric field} \label{Fig.5b}
\end{figure}

The rotation frequency is
\begin{equation}
\omega_{E}=\frac{\alpha_{T}E^{2}}{\hbar}
\label{2.32}
\end{equation}

Besides the rotations the transitions from the vector polarization
into the tensor one and spin dichroism appear.
Moreover, the spin dichroism leads to the appearance of the tensor
polarization.

 Let us compare the frequency and the angle of polarization vector
rotation caused by the EDM with those caused by the birefringence
effect.

1. The spin rotation frequency caused by the EDM is determined by
the formula (\ref{2.3})
\begin{equation}
\omega_{edm}=\frac{d\,E}{\hbar}+\frac{d}{\hbar}\beta B.
\label{2.33}
\end{equation}
We can get  $\omega_{edm}\approx 3\cdot10^{-7}rad/s$ for the
storage ring with $E=3.5\,MV/m$, $B=0.2\,T$ and expected value of
the deuteron EDM $d\sim10^{-27}e\cdot cm$ and $\omega_{edm}\approx
3\cdot10^{-9}rad/s$ for EDM $d\sim10^{-29}e\cdot cm$.

2. The spin rotation frequency caused by the phenomena of
birefringence in a residual gas:
\begin{equation}
\omega=\frac{2\pi N}{M \gamma}\,\hbar \,\texttt{Re}d_{1}.
\label{2.34}
\end{equation}
Using the last formula one can get
$\omega\approx2\cdot10^{-7}\,rad/s$ for $N=10^{9}\,cm^{-3}$
(suppose the pressure inside the storage ring $\sim 10^{-7}$
Torr), $\texttt{Re}d_{1}\sim10^{-13}$. This effect depends on the
density $N$ (depends on the pressure inside the storage ring).

3. The spin rotation frequency caused by the phenomenon of
birefringence in the gas jet (gas target), which is used for the
beam extraction to the polarimeter \cite{project}
(Fig.\ref{Polarimeter}):

\begin{figure}
\epsfxsize = 8 cm \centerline{\epsfbox{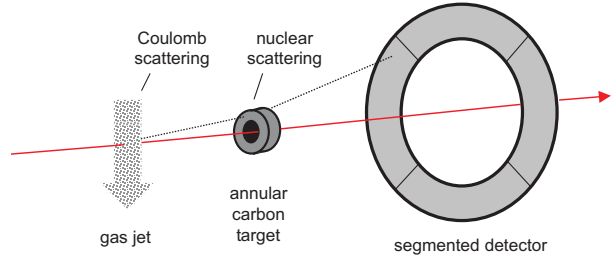}}
\caption{Polarimeter \cite{project}} \label{Polarimeter}
\end{figure}
 In this case  the effective rotation  frequency (or the rotation angle
for $\tau=1s$) is determined as follows:
\begin{equation}
\omega_{t\,eff}\equiv\varphi_{t}=\frac{2\pi
N_{t}}{k}\,l\,\texttt{Re}d_{1}\,\nu,
\label{2.35}
\end{equation}
where  $\omega_{t\,eff}$ is the effective frequency of spin
rotation , $\varphi_{t}$ is the deuteron spin rotation angle for
$\tau=1s$,  $N_{t}$ is the density of the target, $l$ is the
length of the target, $\nu$ is the frequency of beam rotation in
the storage ring, $k=\frac{M \gamma v}{\hbar}$ is the deuteron
wave number.

 Really, the frequency $\omega_{t}$ of the spin rotation in a
 matter is
\begin{equation}
\omega_{t}=\frac{2\pi N_{t}}{M
\gamma}\,\hbar\,\texttt{Re}d_{1}
\label{2.36}
\end{equation}
The spin rotation angle $\theta_{\tau}=\omega_t \tau_t$, where
$\tau_t=\frac{l}{v}$ is the deuteron flying time in the target.
The angle of rotation at 1 second is $\theta=\omega_t \tau_t \nu$
(during a second the deuteron passes through the gas target $\nu$
times). As a result,
\begin{equation}
\omega_{t\,eff}=\omega_{t}\,\tau_{t}\,\nu,
\label{2.37}
\end{equation}
so one can get
\begin{equation}
\omega_{t\,eff}=\frac{2\pi N_t}{M
\gamma}\,\hbar\,\texttt{Re}d_{1}\frac{l}{v}\,\nu=\frac{2\pi
j}{k}\,\texttt{Re}d_{1} \nu,
\label{2.38}
\end{equation}
where $j=N_t\,l$. Then using the experimental parameters
\cite{project} $j=10^{15}\,cm^{-2}$, $\nu\approx 10^{5}-10^{6}$ we
have $\omega_{t\,eff}\approx 10^{-5}-10^{-4}\,rad/s$.

So the angle of polarization vector rotation  for $\tau=1s$ caused
by the phenomenon of birefringence in the gas jet of polarimeter
appears by two orders of magnitude greater than the angle of
rotation due to the EDM. The additional contribution to spin
rotation is also provided by the solid carbon target of
polarimeter (see Figure.\ref{Polarimeter}).

4. The estimation for the spin rotation frequency caused by the
birefringence in an electric field according to (\ref{2.30}) for
$\alpha_{T}\sim10^{-37}cm^{3}$ and $E=3.5MV/m$ is
$\omega_{E}\sim10^{-6}rad/s$.

Let us estimate the value of spin dichroism. This characteristic
is given by the parameter $\chi=-vN(\sigma_1-\sigma_0)$ (see
expressions (\ref{2.25},\ref{2.26})):

- in the case of the scattering by the residual gas we have
$\left| \chi \right| \sim 0.5\cdot10^{-6}$ for $N \sim 10^{9}$
cm$^{-3}$;

- if a deuteron passes through the gas jet (gas target) for
$\tau=1$s is
\begin{equation}
\chi_{t}={j}(\sigma_{1}-\sigma_{0})\,\nu,
\label{2.40}
\end{equation}
then for $j=10^{15}\,cm^{-2}$ and $\nu\approx6\cdot10^{5}$ one can
get $\chi_{t}\sim 3.4\cdot10^{-5}$. So we can conclude that there
is the significant beam spin dichroism.

\section{CONCLUSION}

 The above analysis shows that the phenomena of deuteron
 birefringence in a matter and an electric field should be
 accurately considered when preparing experiments for the EDM
 search with a storage ring, because they could imitate the spin
 rotation due to the EDM. Moreover, study of these effects in such
 experiments could
 provide to measure both the spin-dependent part of the amplitude of
 the coherent elastic scattering of a deuteron by a nucleus at
 the zero angle and the tensor electric polarizability of a deuteron.

 It should be also mentioned that if the nuclei in the gas jet are
 polarized, then according to \cite{4} the P-,T-odd spin rotation and
 dichroism appear in the storage ring. They are caused by the T-odd
 nucleon-nucleon
 interaction of a deuteron with a polarized nucleus, in particular, interaction
 described as $V_{P,T} \sim \vec{S} \left[ \vec{p}_N \times \vec{n}
 \right]$, where $\vec{p}_N$ is the polarization vector of gas
 target.

  P-even T-odd spin rotation and dichroism of deuterons (nuclei)
  caused by the interaction either $V_T \sim (\vec{S} [ \vec{p}_N \times \vec{n}
  ])(\vec{S}\vec{n})$ or
  $V^{\prime}_T \sim Sp \rho_J [([\vec{S} \times \vec{n}] \vec{J}) (\vec{J} \vec{n})]$
  also could be observed
  \cite{4} (here $J \ge 1$ is the spin of the polarized target
  nuclei, $\rho_J$ is the spin matrix density of the target
  nuclei).

\end{document}